\documentclass[prc,aps,amsfonts,amsmath,superscriptaddress,floatfix,preprint]{revtex4}
\usepackage{graphicx} 
\usepackage{bm}% bold math
\usepackage{braket}
\usepackage[colorlinks = true,
linkcolor = blue,
urlcolor  = blue,
citecolor = blue,
anchorcolor = blue]{hyperref}
\begin{document}
\title{Nuclear energy density functionals from empirical ground-state densities}
\author{Giacomo Accorto}
\affiliation{Department of Physics, Faculty of Science, University of
  Zagreb, HR-10000 Zagreb, Croatia}
\author{Tomoya Naito}
\affiliation{Department of Physics, Graduate School of Science,
  The University of Tokyo, Tokyo 113-0033, Japan}
\affiliation{RIKEN Nishina Center, Wako 351-0198, Wako 351-0198, Japan}
\author{Haozhao Liang}
\affiliation{Department of Physics, Graduate School of Science,
  The University of Tokyo, Tokyo 113-0033, Japan}
\affiliation{RIKEN Nishina Center, Wako 351-0198, Wako 351-0198, Japan}
\author{Tamara Nik\v{s}i\'{c}}
\affiliation{Department of Physics, Faculty of Science, University of
  Zagreb, HR-10000 Zagreb, Croatia}
\author{Dario Vretenar}
\affiliation{Department of Physics, Faculty of Science, University of
  Zagreb, HR-10000 Zagreb, Croatia}
\affiliation{State Key Laboratory of Nuclear Physics and Technology,
  School of Physics, Peking University, Beijing 100871, China}
\begin{abstract}
  A model is developed, based on the density functional perturbation theory and the inverse Kohn-Sham method, that can be used to improve relativistic nuclear energy density functionals towards an exact but unknown Kohn-Sham exchange-correlation functional.
  The improved functional is determined by empirical exact ground-state densities of finite systems.
  A test of the model and an illustrative calculation are performed, starting from two different approximate functionals, to reproduce the parameters and density dependence of a target functional, using exact ground-state densities of symmetric $ N = Z $ systems.
\end{abstract}
\date{\today}
\maketitle
% -------------------------------------------------------------------------------------------------------------------
% 
\section{Introduction}
\par
Nuclear energy density functionals (EDFs) have been developed, more or less systematically, over the last two decades into a universal theoretical framework for the analysis of ground-state properties,  low-energy collective excitations, and reaction dynamics of medium-heavy and heavy nuclei \cite{
  Nakatsukasa2016Rev.Mod.Phys.88_045004,
  Schunck_2019}.
At present no other theoretical approach can be used to consistently describe diverse emergent nuclear phenomena with the same level of simplicity and accuracy, and at a comparable computational cost.
Based on universal EDFs, a number of microscopic models,
such as
the random phase approximation \cite{
  Nakatsukasa2007Phys.Rev.C76_024318,
  Liang2013Phys.Rev.C87_054310,
  Niksic2013Phys.Rev.C88_044327},
the interacting boson model \cite{
  Arima1975Phys.Rev.Lett.35_1069,
  Nomura2020Phys.Rev.C102_054313},
and the generator coordinate method \cite{
  Hill1953Phys.Rev.89_1102,
  Griffin1957Phys.Rev.108_311,
  Zhao2020Phys.Rev.C101_064605},
have also been designed to explore low-energy spectroscopy and large-amplitude dynamics on a quantitative level, and to calculate various details for astrophysical applications.  
\par
The exact but unknown nuclear EDF must be approximated by functionals of powers and gradients of 
ground-state nucleon densities and currents.
Even though a great number of studies have been devoted to the microscopic formulation of a universal EDF framework (for a recent review, see Ref.~\cite{Furnstahl_2020}), the most successful nuclear EDFs are either semi-empirical or completely phenomenological.
In a semi-empirical approach, one starts from a microscopically motivated \textit{ansatz} for the nucleonic density dependence of the energy of a system of protons and neutrons.
Part of the parameters of such a functional can be determined, at least qualitatively, by microscopic calculations of the energy of isospin symmetric and asymmetric infinite nuclear matter as a function of the nucleonic density (or Fermi momentum).
The remaining parameters are usually adjusted to selected ground-state data,
e.g., masses and charge radii, of an arbitrarily large set of nuclei.
Fully phenomenological EDFs, for instance Skyrme, Gogny, and relativistic ones, usually take into account some empirical properties of nuclear matter at saturation,
but all parameters are adjusted to ground-state data of finite nuclei. 
\par
The question we address in this work is how to improve a given functional towards the exact but unknown nuclear EDF.
One could start, for instance, from a general expansion in powers of densities and currents and retain all terms allowed by symmetries up to a given order.
Such a functional could be derived, in principle, from a microscopic theory (low-energy QCD) that describes the underlying many-body dynamics.
The problem, however,  is that available low-energy nuclear data can only constrain a relatively small subset of terms and determine the corresponding parameters.
Moreover, nuclear EDFs are ``sloppy'', that is, they generally exhibit an \textit{exponential} range of sensitivity to parameter variations, and one finds many \textit{soft} linear combinations of bare model parameters that are poorly constrained by data.
This often indicates the presence of low-dimensional effective functionals associated with the relevant (\textit{stiff}) parameter combinations.
In Refs.~\cite{NV_2016,NIV_2017}, we considered, in the context of nuclear structure, the interesting problem of a systematic construction of reduced low-dimensional functionals from a more complete but sloppy framework.
Using methods of information geometry, it has been shown how to systematically construct effective EDFs of successively 
lower dimension in parameter space until sloppiness is eventually eliminated and the resulting functional contains only stiff combinations of parameters.
\par
Instead of using low-energy data to reduce the complexity of a very general functional, one could also start from a relatively simple functional form and improve it towards the exact but unknown EDF.
Such an expansion must, of course, be constrained by available data.
In the spirit of density functional theory (DFT) \cite{
  Hohenberg1964Phys.Rev.136_B864,
  Kohn1965Phys.Rev.140_A1133,
  Kohn1999Rev.Mod.Phys.71_1253},
the empirical (exact) ground-state densities should determine the improved EDF.
In fact, the inverse problem of DFT is formulated as a density-to-potential inversion that, starting from a given exact ground-state density, determines the true Kohn-Sham (KS) exchange-correlation potential \cite{Wang_1993,Zhao1993J.Chem.Phys.98_543,Leeuwen_1994,Zhao1994Phys.Rev.A50_2138,Jensen.18,KSH_2019,Kanungo_2019,Nam_2020,Garrigue_2021}.
The first application of the inverse KS method to nuclear EDFs has recently been reported in Ref.~\cite{Accorto.20}.
The inverse Kohn-Sham (IKS) method is often used to benchmark and test the accuracy of various approximate exchange-correlation functionals but, since its implementation depends on the exact density of a specific system, it does not provide direct information on the universal EDF.
\par
The problem of improving the functional starting from exact ground-state densities has recently been  addressed in Ref.~\cite{Naito.19}, where a model was introduced based on a combination of the IKS method and the density functional perturbation theory (DFPT).
The idea is to consider the difference, supposedly small, between the known functional and the exact but unknown EDF as a first-order perturbation.
In Ref.~\cite{Naito.19}, the method was successfully tested in benchmark calculations of systems of noble-gas atoms.
\par
In this work, we apply the novel approach to atomic nuclei and, in particular, to relativistic nuclear EDFs.
The principal reason for considering relativistic functionals is the fact that they automatically take into account the spin-orbit potential.
The strong coupling between the orbital angular momentum and nucleon spin accounts for the empirical magic numbers and shell gaps.
While in atomic systems the effect of the spin-orbit coupling is perturbative, in nuclei the energy spacings between spin-orbit partner single-nucleon states can be as large as the gaps between major shells.
However, because the spin-orbit potential is a completely phenomenological addition to the non-relativistic KS potential, it cannot be determined from the ground-state density using the IKS method \cite{Accorto.20}.
For relativistic EDFs,
the spin-orbit potential emerges automatically with the proper strength as the constructive combination of the scalar and vector nucleon self-energies. 
\par
In Sec.~\ref{sec:II}, we introduce the formalism of DFPT and IKS, and develop the model that will be used in this work.
In Sec.~\ref{sec:test}, a test case is discussed, and Sec.~\ref{sec:example} presents an illustrative calculation.
Section \ref{sec:conclusion} contains a brief summary and outlook for future studies. 
% -------------------------------------------------------------------------------------------------------------------
% 
\section{Nuclear energy density functionals improved by exact ground-state densities}
\label{sec:II}
\subsection{Density functional perturbation theory}
\label{subsec:DFPT}
\par
Here, we apply the method of Ref.~\cite{Naito.19} to improve a relativistic nuclear EDF, starting from a given empirical ground-state density.
It is assumed that the unknown 
exact Hartree-exchange-correlation (interaction) functional can be written in the following form:
\begin{equation}
  E_{\text{int}}
  \left[ \rho \right]
  =
  E_{\text{int}}^{(0)} \left[ \rho \right]
  +
  \lambda E_{\text{int}}^{(1)} \left[ \rho \right]
  +
  \mathcal{O} \left( \lambda^2 \right),
\end{equation}
where $E_{\text{int}}^{(0)} \left[ \rho \right]$ denotes the known functional that we wish to make better and $\lambda$ is a small parameter.
The main premise of this method, therefore, is that 
the difference between the exact functional $E_{\text{int}} \left[ \rho \right] $ and the starting functional $E_{\text{int}}^{(0)} \left[ \rho \right]$ is small
enough to be treated perturbatively.
The correction will be determined by the exact ground-state density, using the IKS approach. 
\par
The Dirac KS single-nucleon equation, derived from the relativistic EDF, reads
\begin{equation}
  \label{Dirac-eq-DFPT}
  \left[
    \bm{\alpha} \cdot \bm{p}
    +
    \beta \left( m + S \left( \bm{r} \right) \right)
    +
    V \left( \bm{r} \right) 
  \right]
  \psi_j \left( \bm{r} \right)
  =
  \epsilon_j \psi_j \left( \bm{r} \right),
\end{equation}
with the scalar and vector KS potentials, respectively,
\begin{equation}
  S \left( \bm{r} \right)
  =
  \left.
    \frac{\delta E \left[ \rho_{\text{V}}, \rho_{\text{S}} \right]}{\delta \rho_{\text{S}}}
    % \right|_{\rho_{\text{V,gs}}, \rho_{\text{S,gs}}}
  \right|_{\text{gs}}
  \quad \text{and} \quad
  V \left( \bm{r} \right)
  =
  \left.
    \frac{\delta E \left[ \rho_{\text{V}}, \rho_{\text{S}} \right]}{\delta \rho_{\text{V}}}
  % \right|_{\rho_{\text{V,gs}}, \rho_{\text{S,gs}}}.
  \right|_{\text{gs}}.
\end{equation}
The corresponding scalar and vector densities
\begin{equation}
  \label{eq:density}
  \rho_{\text{S,gs}} \left( \bm{r} \right)
  =
  \sum_{j \in \text{occ}}
  \psi_j^\dagger \left( \bm{r} \right)
  \beta \psi_j \left( \bm{r} \right)
  \quad \text{and} \quad
  \rho_{\text{V,gs}} \left( \bm{r} \right)
  =
  \sum_{j \in \text{occ}}
  \psi_j^\dagger \left( \bm{r} \right)
  \psi_j \left( \bm{r} \right)
\end{equation}
are obtained from the self-consistent solutions of the single-nucleon Dirac KS equation
in the \textit{no-sea} approximation,
which omits explicit contributions of negative-energy states to densities and currents
\cite{Ser.86,Rei.89,Ring.96} and, thus, the sums run only over
the occupied positive-energy single-nucleon orbitals.  
\par
The exact Dirac spinors, that is, the solutions of Eq.~(\ref{Dirac-eq-DFPT}) for the exact EDF, can also be expanded to the first order in $\lambda$: 
\begin{equation}
  \label{eq:psi-expansion}
  \psi_j \left( \bm{r} \right)
  =
  \psi^{(0)}_j \left( \bm{r} \right)
  +
  \lambda \psi^{(1)}_j \left( \bm{r} \right)
  +
  \mathcal{O} \left( \lambda^2 \right),
\end{equation}
where the first-order correction is orthogonal to the zeroth-order spinor 
\begin{equation}
  \label{eq:orthogonality}
  \int
  \psi^{(0)\dagger}_j \left( \bm{r} \right)
  \psi^{(1)}_j \left( \bm{r} \right)
  \, d\bm{r}
  =
  0.
\end{equation}
The exact ground-state scalar and vector densities
\begin{subequations}
  \begin{align}
    \rho_{\text{S,gs}} \left( \bm{r} \right)
    = & \, 
        \sum_{j \in \text{occ}}
        \psi_j^{(0)\dagger} \left( \bm{r} \right)
        \beta
        \psi_j^{(0)} \left( \bm{r} \right)
        \notag \\
    & \,
      +
      \lambda
      \left(
      \sum_{j \in \text{occ}}
      \psi_j^{(1)\dagger} \left( \bm{r} \right)
      \beta
      \psi_j^{(0)} \left( \bm{r} \right) 
      +
      \sum_{j \in \text{occ}}
      \psi_j^{(0)\dagger} \left( \bm{r} \right) \beta \psi_j^{(1)} \left( \bm{r} \right)
      \right)
      +
      \mathcal{O} \left( \lambda^2 \right),
      \label{rho_s} \\
    %%%
    \rho_{\text{V,gs}} \left( \bm{r} \right)
    = & \,
        \sum_{j \in \text{occ}}
        \psi_j^{(0)\dagger} \left( \bm{r} \right)
        \psi_j^{(0)} \left( \bm{r} \right)
        \notag \\
      & \, 
        +
        \lambda
        \left(
        \sum_{j \in \text{occ}}
        \psi_j^{(1)\dagger} \left( \bm{r} \right)
        \psi_j^{(0)} \left( \bm{r} \right)
        +
        \sum_{j \in \text{occ}}
        \psi_j^{(0)\dagger} \left( \bm{r} \right)
        \psi_j^{(1)} \left( \bm{r} \right)
        \right)
        +
        \mathcal{O} \left( \lambda^2 \right), 
        \label{rho_v} 
  \end{align}
\end{subequations}
take the form:
\begin{equation}
  \label{dens_exp}
  \rho_{\text{V(S),gs}} \left( \bm{r} \right)
  =
  \rho_{\text{V(S)}}^{(0)} \left( \bm{r} \right)
  +
  \lambda
  \rho_{\text{V(S)}}^{(1)} \left( \bm{r} \right)
  +
  \mathcal{O} \left( \lambda^2 \right),
\end{equation}
where the first term on the right-hand side denotes the densities derived by the known functional.
We will assume that, given the exact densities $\rho_{\text{V(S),gs}} \left( \bm{r} \right) $,
one can use the IKS method to calculate the \textit{exact} single-nucleon Dirac spinors $\psi_j \left( \bm{r} \right) $ and energies $\epsilon_j$. 
\par
The ground-state energy can also be decomposed as follows:
\begin{equation}
  \label{E_gs}
  % E_{\text{gs}} = E_{kin}+E_{ext} + E_{\text{int}}^{(0)}+ \lambda E_{\text{int}}^{(1)}.
  E_{\text{gs}}
  =
  E_{\text{kin}} % \left[ \rho_{\text{gs}} \right]
  +
  E_{\text{int}}^{(0)} \left[ \rho_{\text{V,gs}}, \rho_{\text{S,gs}}\right]
  +
  \lambda E_{\text{int}}^{(1)} \left[ \rho_{\text{V,gs}}, \rho_{\text{S,gs}} \right],
\end{equation} 
where the KS kinetic energy reads
\begin{equation}
  E_{\text{kin}} % \left[ \rho_{\text{gs}} \right]
  =
  \sum_{j \in \text{occ}}
  \int
  \psi_j^\dagger \left( \bm{r} \right)
  \hat{t}
  \psi_j \left( \bm{r} \right)
  \, d\bm{r}  ,
\end{equation}
and $\hat{t}=\bm{\alpha} \cdot \bm{p} + \beta m$.
By expanding the
Dirac spinors as in Eq.~(\ref{eq:psi-expansion})
% $\psi \left( \bm{r} \right)  = \psi^{(0)} \left( \bm{r} \right) +\lambda \psi^{(1)} \left( \bm{r} \right) +\mathcal{O}(\lambda^2)$ 
and retaining only terms linear in $\lambda$, we obtain
the following expression for the kinetic energy
\begin{align}
  E_{\text{kin}} % \left[ \rho_{\text{gs}} \right]
  = & \, 
    \sum_{j \in \text{occ}}
    \int
    \psi_j^{(0)\dagger} \left( \bm{r} \right)
    \hat{t}
    \psi_j^{(0)} \left( \bm{r} \right)
    \, d\bm{r}
    \notag \\
    & \,
      +
      \lambda
      \sum_{j \in \text{occ}}
      \int
      \psi_j^{(1)\dagger} \left( \bm{r} \right)
      \hat{t}
      \psi_j^{(0)} \left( \bm{r} \right)
      \, d\bm{r} 
      +
      \lambda
      \sum_{j \in \text{occ}}
      \int
      \psi_j^{(0)\dagger} \left( \bm{r} \right)
      \hat{t}
      \psi_j^{(1)} \left( \bm{r} \right)
      \, d\bm{r}
      +
      \mathcal{O} \left( \lambda^2 \right).
      \label{E_kin}
\end{align}
% The energy induced by the external potential $V_{ext} \left( \bm{r} \right) $ can also be expanded
% \begin{equation}
%   E_{ext} = \int{V_{ext} \left( \bm{r} \right) \rho^{(0)}_{V,gs} \left( \bm{r} \right) d\bm{r}} + \lambda  \int{V_{ext} \left( \bm{r} \right) \rho^{(1)}_{V,gs} \left( \bm{r} \right) d\bm{r}} + \mathcal{O}(\lambda^2).
% \end{equation}
% The interaction energy reads
% \begin{equation}
%   \%label{eq:E_int}
%   E_{\text{int}}[\rho_{\text{V}},\rho_{\text{S}}] = \frac{1}{2} \int{S_{KS} \left( \bm{r} \right) \rho_{S,gs} \left( \bm{r} \right) d\bm{r}} + \frac{1}{2} \int{V_{KS} \left( \bm{r} \right) \rho_{V,gs} \left( \bm{r} \right) d\bm{r}},
% \end{equation}
% with scalar and vector potential calculated from the EDF
% \begin{equation}
%   S_{KS} \left( \bm{r} \right)  = \left.\frac{\delta \mathcal{E}_{\text{int}}[\rho_{\text{V}},\rho_{\text{S}}]}{\delta \rho_{\text{S}}}\right|_{\rho_{V,gs},\rho_{S,gs}} \quad \textnormal{and}  \quad
%   V_{KS} \left( \bm{r} \right)  = \left.\frac{\delta \mathcal{E}_{\text{int}}[\rho_{\text{V}},\rho_{\text{S}}]}{\delta \rho_{\text{V}}}\right|_{\rho_{V,gs},\rho_{S,gs}} .
% \end{equation}
The second and third terms on the right-hand side of Eq.~(\ref{E_gs}) denote the interaction (Hartree-exchange-correlation) contribution to the total energy.
Using the expansion [Eq.~(\ref{dens_exp})] for 
the ground-state densities up to terms linear in $\lambda$, one derives:
\begin{subequations}
  \begin{align}
    E_{\text{int}}^{(0)} \left[ \rho_{\text{V,gs}}, \rho_{\text{S,gs}} \right]
    = & \, 
        E_{\text{int}}^{(0)} \left[ \rho_{\text{V,gs}}^{(0)}, \rho_{\text{S,gs}}^{(0)} \right]
        +
        \lambda
        \int
        \left.
        \frac{\delta E_{\text{int}}^{(0)} \left[ \rho_{\text{V}}, \rho_{\text{S}} \right]}{\delta \rho_{\text{V}}}
        % \right|_{\rho^{(0)}_{\text{V,gs}},\rho^{(0)}_{\text{S,gs}}}
        \right|_{\text{gs}^{(0)}}
        \rho_{\text{V,gs}}^{(1)} \left( \bm{r} \right)
        \, d \bm{r}
        \notag \\
      & \,
        +
        \lambda
        \int
        \left.
        \frac{\delta E_{\text{int}}^{(0)} \left[ \rho_{\text{V}}, \rho_{\text{S}} \right]}{\delta \rho_{\text{S}}}
        % \right|_{\rho^{(0)}_{\text{V,gs}},\rho^{(0)}_{\text{S,gs}}}
        \right|_{\text{gs}^{(0)}}
        \rho_{\text{S,gs}}^{(1)} \left( \bm{r} \right)
        \, d\bm{r}
        +
        \mathcal{O} \left( \lambda^2 \right), 
        \label{eq:E(0)_int} \\
    \lambda
    E_{\text{int}}^{(1)} \left[ \rho_{\text{V,gs}},\rho_{\text{S,gs}} \right]
    = & \, 
        \lambda E_{\text{int}}^{(1)} \left [\rho_{\text{V,gs}}^{(0)},\rho_{\text{S,gs}}^{(0)} \right]
        +
        \mathcal{O} \left( \lambda^2 \right).
        \label{eq:E(1)_int}
  \end{align}
\end{subequations}
% By using the expansions for the Dirac spinors, the vector and scalar density can be written as:
% \begin{align}
%   \rho_{V,gs} \left( \bm{r} \right)  &= \sum_{j \in \text{occ}}{\psi_j^{(0)\dagger} \left( \bm{r} \right)  \psi_j^{(0)} \left( \bm{r} \right) } + \lambda  \sum_{j \in \text{occ}}{\psi_j^{(1)\dagger} \left( \bm{r} \right)  \psi_j^{(0)} \left( \bm{r} \right) } 
%   +  \sum_{j \in \text{occ}}{\psi_j^{(0)\dagger} \left( \bm{r} \right)  \psi_j^{(1)} \left( \bm{r} \right) } +  \mathcal{O}(\lambda^2),\\
    %     \rho_{S,gs} \left( \bm{r} \right)  &= \sum_{j \in \text{occ}}{\psi_j^{(0)\dagger} \left( \bm{r} \right)  \beta \psi_j^{(0)} \left( \bm{r} \right) } 
    %     + \lambda  \sum_{j \in \text{occ}}{\psi_j^{(1)\dagger} \left( \bm{r} \right) \beta \psi_j^{(0)} \left( \bm{r} \right) } 
    %     +  \sum_{j \in \text{occ}}{\psi_j^{(0)\dagger} \left( \bm{r} \right) \beta \psi_j^{(1)} \left( \bm{r} \right) } +  \mathcal{O}(\lambda^2).
    %   \end{align}
Inserting the expressions Eqs.~(\ref{rho_v}) and (\ref{rho_s}) for the 
first-order density corrections $\rho_{\text{V,gs}}^{(1)}$ and $\rho_{\text{S,gs}}^{(1)}$, respectively, into the expansion for the interaction energy Eqs.~(\ref{eq:E(0)_int}) and (\ref{eq:E(1)_int}),
together with 
the expression for the kinetic energy Eq.~(\ref{E_kin}), the ground-state energy reads
\begin{align}
  E_{\text{gs}}
  &=  \,
      E^{(0)}_{\text{kin}}
      +
      E^{(0)}_{\text{int}} \left[ \rho_{\text{V,gs}}^{(0)},\rho_{\text{S,gs}}^{(0)} \right]
      \notag \\
    & \,
      +
      \lambda
      \sum_{j \in \text{occ}}
      \int
      \psi_j^{(1)\dagger} \left( \bm{r} \right)
      \left[ \hat{t} 
      +
      \left.
      \frac{\delta E_{\text{int}}^{(0)} \left[ \rho_{\text{V}}, \rho_{\text{S}} \right]}{\delta \rho_{\text{V}}}
      % \right|_{\rho^{(0)}_{\text{V,gs}}, \rho^{(0)}_{\text{S,gs}}}
      \right|_{\text{gs}^{(0)}}
      +
      \beta
      \left.
      \frac{\delta E_{\text{int}}^{(0)} \left[ \rho_{\text{V}}, \rho_{\text{S}} \right]}{\delta \rho_{\text{S}}}
      % \right|_{\rho^{(0)}_{\text{V,gs}},\rho^{(0)}_{\text{S,gs}}}
      \right|_{\text{gs}^{(0)}}
      \right]
      \psi_j^{(0)} \left( \bm{r} \right)
      \, d \bm{r}
      \notag \\
    & \,
      +
      \lambda
      \sum_{j \in \text{occ}}
      \int
      \psi_j^{(0)\dagger} \left( \bm{r} \right)
      \left[
      \hat{t}  
      +
      \left.
      \frac{\delta E_{\text{int}}^{(0)} \left[ \rho_{\text{V}}, \rho_{\text{S}} \right]}{\delta \rho_{\text{V}}}
      % \right|_{\rho^{(0)}_{\text{V,gs}},\rho^{(0)}_{\text{S,gs}}}
      \right|_{\text{gs}^{(0)}}
      +
      \beta
      \left.
      \frac{\delta E_{\text{int}}^{(0)} \left[ \rho_{\text{V}}, \rho_{\text{S}} \right]}{\delta \rho_{\text{S}}}
      % \right|_{\rho^{(0)}_{\text{V,gs}},\rho^{(0)}_{\text{S,gs}}}
      \right|_{\text{gs}^{(0)}}
      \right]
      \psi_j^{(1)}
      \left( \bm{r} \right)
      \, d\bm{r}
      \notag \\
    & \,
      +
      \lambda
      E_{\text{int}}^{(1)} \left[ \rho_{\text{V,gs}}^{(0)},\rho_{\text{S,gs}}^{(0)} \right]
      +
      \mathcal{O} \left( \lambda^2 \right).
      \label{eq:Egs}
\end{align}
One notices that the expression in square brackets represent the 
zeroth-order (unperturbed) Dirac Hamiltonian, that is 
\begin{equation}
  \left[
    \hat{t}
    +
    \left.
      \frac{\delta E_{\text{int}}^{(0)} \left[ \rho_{\text{V}}, \rho_{\text{S}} \right]}{\delta \rho_{\text{V}}}
    % \right|_{\rho^{(0)}_{\text{V,gs}},\rho^{(0)}_{\text{S,gs}}}
    \right|_{\text{gs}^{(0)}}
    +
    \beta
    \left.
      \frac{\delta E_{\text{int}}^{(0)} \left[ \rho_{\text{V}}, \rho_{\text{S}} \right]}{\delta \rho_{\text{S}}}
    % \right|_{\rho^{(0)}_{\text{V,gs}},\rho^{(0)}_{\text{S,gs}}}
    \right|_{\text{gs}^{(0)}}
  \right]
  \psi_j^{(0)}
  =
  \epsilon_j^{(0)} \psi_j^{(0)}.
\end{equation}
The corresponding terms in Eq.~(\ref{eq:Egs}) vanish because of the orthogonality relation [Eq.~(\ref{eq:orthogonality})] and,
thus, a much simpler relation for the ground-state energy reads % is obtained:
\begin{equation}
  \label{eq:E-gs1}
  E_{\text{gs}}
  =
  E^{(0)}_{\text{kin}}
  +
  E^{(0)}_{\text{int}} \left[ \rho_{\text{V,gs}}^{(0)},\rho_{\text{S,gs}}^{(0)} \right]
  +
  \lambda E_{\text{int}}^{(1)} \left[ \rho_{\text{V,gs}}^{(0)},\rho_{\text{S,gs}}^{(0)} \right]  
  +
  \mathcal{O} \left( \lambda^2 \right).
\end{equation}
On the other hand, the ground state energy
$E_{\text{gs}} = E_{\text{kin}} % \left[ \rho_{\text{gs}} \right]
+ E_{\text{int}} \left[ \rho_{\text{gs}} \right]$, 
can be written in the following form 
\begin{equation}
  E_{\text{gs}}
  =
  \sum_{j \in \text{occ}}
  \epsilon_j
  +
  E_{\text{int}} \left[ \rho_{\text{V,gs}},\rho_{\text{S,gs}} \right]
  -
  \int
  \left.
    \frac{\delta E_{\text{int}} \left[ \rho_{\text{V}}, \rho_{\text{s}} \right]}{\delta \rho_{\text{V}}}
  \right|_{\text{gs}}
  \rho_{\text{V,gs}} \left( \bm{r} \right)
  \, d\bm{r} 
  -
  \int
  \left.
    \frac{\delta E_{\text{int}} \left[ \rho_{\text{V}}, \rho_{\text{s}} \right]}{\delta \rho_{\text{S}}}
  \right|_{\text{gs}}
  \rho_{\text{S,gs}} \left( \bm{r} \right)
  \, d\bm{r} ,
\end{equation}
where the Dirac KS equation has been used to eliminate the explicit contribution of the kinetic energy term, and $\epsilon_j$ are the exact single-particle energies with the summation over occupied states.
If we separate the zeroth-order and first-order terms of the exact interaction functional $E_{\text{int}} \left[ \rho_{\text{gs}} \right]$, then 
\begin{align}
  E_{\text{gs}}
  = & \,
      \sum_{j \in \text{occ}}
      \epsilon_j
      +
      E^{(0)}_{\text{int}} \left[ \rho_{\text{V,gs}},\rho_{\text{S,gs}} \right] 
      -
      \int
      \left.
      \frac{\delta E^{(0)}_{\text{int}} \left[ \rho_{\text{V}}, \rho_{\text{s}} \right]}{\delta \rho_{\text{V}}}
      \right|_{\text{gs}}
      \rho_{\text{V,gs}} \left( \bm{r} \right)
      \, d\bm{r} 
      -
      \int
      \left.
      \frac{\delta E^{(0)}_{\text{int}} \left[ \rho_{\text{V}}, \rho_{\text{s}} \right]}{\delta \rho_{\text{S}}}
      \right|_{\text{gs}}
      \rho_{\text{S,gs}} \left( \bm{r} \right)
      \, d\bm{r}
      \notag \\
    & \,
      +
      \lambda
      E^{(1)}_{\text{int}} \left[ \rho_{\text{V,gs}},\rho_{\text{S,gs}} \right]  
      -
      \lambda
      \int
      \left.
      \frac{\delta E^{(1)}_{\text{int}} \left[ \rho_{\text{V}}, \rho_{\text{s}} \right]}{\delta \rho_{\text{V}}}
      \right|_{\text{gs}}
      \rho_{\text{V,gs}} \left( \bm{r} \right)
      \, d\bm{r} 
      -
      \lambda
      \int
      \left.
      \frac{\delta E^{(1)}_{\text{int}} \left[ \rho_{\text{V}}, \rho_{\text{s}} \right]}{\delta \rho_{\text{S}}}
      \right|_{\text{gs}}
      \rho_{\text{S,gs}} \left( \bm{r} \right)
      \, d\bm{r}.
      \label{eq:E-gs2}
\end{align}
From Eq.~(\ref{eq:E-gs1}), we can express the first-order correction to the interaction energy as a function of the zeroth-order ground-state densities
\begin{equation}
  \lambda
  E_{\text{int}}^{(1)} \left[ \rho_{\text{V,gs}}^{(0)},\rho_{\text{S,gs}}^{(0)} \right]
  =
  E_{\text{gs}}
  -
  E^{(0)}_{\text{kin}}
  -
  E^{(0)}_{\text{int}} \left[ \rho_{\text{V,gs}}^{(0)},\rho_{\text{S,gs}}^{(0)} \right]
  =
  E_{\text{gs}}  - E_{\text{gs}}^{(0)}, 
\end{equation}
where obviously $E_{\text{gs}}^{(0)}$ denotes the ground-state energy calculated with the known functional.
\par
Next, Eq.~(\ref{eq:E-gs2}) for $E_{\text{gs}}$ is inserted in this expression and the following relation is obtained,
\begin{align}
  & \lambda E_{\text{int}}^{(1)} \left[ \rho_{\text{V,gs}}^{(0)},\rho_{\text{S,gs}}^{(0)} \right]
    \notag \\
    = & \, 
    \sum_{j \in \text{occ}}
    \epsilon_j
    +
    E^{(0)}_{\text{int}} \left[ \rho_{\text{V,gs}},\rho_{\text{S,gs}} \right]
    -
    \int
    \left.
    \frac{\delta E^{(0)}_{\text{int}} \left[ \rho_{\text{V}}, \rho_{\text{s}} \right]}{\delta \rho_{\text{V}}}
    \right|_{\text{gs}}
    \rho_{\text{V,gs}} \left( \bm{r} \right)
    \, d \bm{r} 
    -
    \int
    \left.
    \frac{\delta E^{(0)}_{\text{int}} \left[ \rho_{\text{V}}, \rho_{\text{s}} \right]}{\delta \rho_{\text{S}}}
    \right|_{\text{gs}}
    \rho_{\text{S,gs}} \left( \bm{r} \right)
    \, d \bm{r}
    \notag \\
  & +
    \lambda
    E^{(1)}_{\text{int}} \left[ \rho_{\text{V,gs}},\rho_{\text{S,gs}} \right] 
    -
    \lambda
    \int
    \left.
    \frac{\delta E^{(1)}_{\text{int}} \left[ \rho_{\text{V}}, \rho_{\text{s}} \right]}{\delta \rho_{\text{V}}}
    \right|_{\text{gs}}\rho_{\text{V,gs}}
    \left( \bm{r} \right)
    \, d\bm{r} 
    -
    \lambda
    \int
    \left.
    \frac{\delta E^{(1)}_{\text{int}} \left[ \rho_{\text{V}}, \rho_{\text{s}} \right]}{\delta \rho_{\text{S}}}
    \right|_{\text{gs}}
    \rho_{\text{S,gs}} \left( \bm{r} \right)
    \, d\bm{r}
    -
    E_{\text{gs}}^{(0)}.
    \label{eq:izraz}
\end{align}
Equation~(\ref{eq:izraz}) is now rearranged so that all terms linear in $\lambda$ (first-order corrections to the interaction functional) are collected on the left-hand side
\begin{align}
  & \lambda
    E_{\text{int}}^{(1)} \left[ \rho_{\text{V,gs}}^{(0)},\rho_{\text{S,gs}}^{(0)} \right] 
    -
    \lambda
    E^{(1)}_{\text{int}} \left[ \rho_{\text{V,gs}},\rho_{\text{S,gs}} \right] 
    +
    \lambda
    \int
    \left.
    \frac{\delta E^{(1)}_{\text{int}} \left[ \rho_{\text{V}}, \rho_{\text{S}} \right]}{\delta \rho_{\text{V}}}
    \right|_{\text{gs}}
    \rho_{\text{V,gs}} \left( \bm{r} \right)
    \, d \bm{r}   
    +
    \lambda
    \int
    \left.
    \frac{\delta E^{(1)}_{\text{int}} \left[ \rho_{\text{V}}, \rho_{\text{S}} \right]}{\delta \rho_{\text{S}}}
    \right|_{\text{gs}}
    \rho_{\text{S,gs}} \left( \bm{r} \right)
    \, d \bm{r}
    \notag \\
  & =
    \sum_{j \in \text{occ}}
    \epsilon_j
    +
    E^{(0)}_{\text{int}} \left[ \rho_{\text{V,gs}},\rho_{\text{S,gs}} \right] 
    -
    \int
    \left.
    \frac{\delta E^{(0)}_{\text{int}} \left[ \rho_{\text{V}}, \rho_{\text{S}} \right]}{\delta \rho_{\text{V}}}
    \right|_{\text{gs}}
    \rho_{\text{V,gs}} \left( \bm{r} \right)
    \, d \bm{r}
    -
    \int
    \left.
    \frac{\delta E^{(0)}_{\text{int}} \left[ \rho_{\text{V}}, \rho_{\text{S}} \right]}{\delta \rho_{\text{S}}}
    \right|_{\text{gs}}
    \rho_{\text{S,gs}} \left( \bm{r} \right)
    \, d \bm{r} 
    -
    E_{\text{gs}}^{(0)}.
    \label{eq:final}
\end{align}
The right-hand side of this equation depends only on the exact ground-state densities and the known functional $E_{\text{int}}^{(0)}$.
For given ground-state empirical densities, therefore, we can calculate all terms on the right-hand side, except the first term which is a sum of the exact single-particle energies.
These energies are, of course, implicit functionals of the exact ground-state densities.
One can, therefore, use the IKS method to calculate the single-particle energies starting from given ground-state densities, a procedure that we describe in the following section.
\par
In practical terms, one must assume a certain \textit{ansatz} for the functional
$E_{\text{int}}^{(1)}\left[ \rho \right]$,
that will also include parameters to be determined from Eq.~(\ref{eq:final}) for a choice of empirical ground-state densities.
There is no guarantee, especially in the case of several undetermined parameters for the first-order correction, that the improved functional will reproduce the exact densities to a desired level of accuracy.
The solution is an iterative procedure \cite{Naito.19}, in which the functional improved in the first iteration step is considered as the known functional for the next iteration, i.e., 
\begin{equation}
  \label{eq:iter}
  E_{\text{int}}^{\text{$ n $-th}} \left[ \rho_{\text{V}}, \rho_{\text{S}} \right]
  =
  E_{\text{int}}^{(0)} \left[ \rho_{\text{V}}, \rho_{\text{S}} \right]
  +
  \sum_{k=1}^{n}
  \lambda E_{\text{int}}^{\text{$(1)$, $ k $-th}} \left[ \rho_{\text{V}}, \rho_{\text{S}} \right],
\end{equation}
and the operation is repeated until the exact densities are reproduced by the solutions of the resulting \textit{$ n $-th} iteration Dirac KS equation to a desired accuracy. 
% 
% ---------------------------------------------------------------------------------------------------------------------------
\subsection{Inverse Kohn-Sham method}
\label{subsec:IKS}
% ---------------------------------------------------------------------------------------------------------------------------
\par
Determining the KS potential for a given density presents an inverse problem.
According to the Hohenberg-Kohn theorem, this inverse problem has a solution and the KS exchange-correlation potential for a given system of interacting particles can be calculated starting from its ground-state density.
Here, we perform the density-to-potential inversion in order to determine the single-particle energies that appear on the right-hand side Eq.~(\ref{eq:final}). 
Starting from the single-nucleon Dirac KS equation~(\ref{Dirac-eq-DFPT}),
% \begin{equation}
%   \label{eq:Dirac1}
%   \left[
%     \bm{\alpha} \cdot \bm{p}
%     +
%     \beta \left( m + S \left( \bm{r} \right) \right)
%     +
%     V \left( \bm{r} \right)
%   \right]
%   \psi_j \left( \bm{r} \right)
%   =
%   E_j \psi_j \left( \bm{r} \right),
% \end{equation}
we rewrite the KS potentials:
$ V_{+} \left( \bm{r} \right) = V \left( \bm{r} \right) + S \left( \bm{r} \right) $
and
$ V_{-} \left( \bm{r} \right) = V \left( \bm{r} \right) - S \left( \bm{r} \right) $,
so that Eq.~(\ref{Dirac-eq-DFPT}) takes the form
\begin{equation}
  \label{eq:Dirac2}
  \left[
    \bm{\alpha} \cdot \bm{p}
    +
    \frac{1}{2}
    \left( \beta - \openone \right)
    \left( m - V_{-} \left( \bm{r} \right) \right)
    +
    \frac{1}{2}
    \left( \beta + \openone \right)
    \left( m + V_{+} \left( \bm{r} \right) \right) 
  \right]
  \psi_j \left( \bm{r} \right) 
  =
  \epsilon_j \psi_j \left( \bm{r} \right).
\end{equation}
By multiplying Eq.~(\ref{eq:Dirac2}) with $\psi_j^\dagger \left( \bm{r} \right)$ from the left and summing over the occupied positive-energy states, one obtains
\begin{align}
  & \sum_{j \in \text{occ}}
  \psi_j^\dagger \left( \bm{r} \right)
  \left( \bm{\alpha} \cdot \bm{p} - \epsilon_j \right)
  \psi_j \left( \bm{r} \right)
  \notag \\
  & +
    \frac{1}{2}
    \left( m - V_{-} \left( \bm{r} \right) \right)
    \sum_{j \in \text{occ}}
    \psi_j^\dagger \left( \bm{r} \right)
    \left( \beta - \openone \right)
    \psi_j \left( \bm{r} \right)
    +
    \frac{1}{2}
    \left( m + V_{+} \left( \bm{r} \right) \right)
    \sum_{j \in \text{occ}}
    \psi_j^\dagger \left( \bm{r} \right)
    \left( \beta + \openone \right)
    \psi_j \left( \bm{r} \right)
    =
    0. 
\end{align}
The scalar and vector densities that appear in this expression
\begin{equation}
  \rho_{\text{S,gs}} \left( \bm{r} \right)
  =
  \sum_{j \in \text{occ}}
  \psi_j^\dagger \left( \bm{r} \right)
  \beta
  \psi_j \left( \bm{r} \right)
  \quad
  \text{and}
  \quad
  \rho_{\text{V,gs}} \left( \bm{r} \right)
  =
  \sum_{j \in \text{occ}}
  \psi_j^\dagger \left( \bm{r} \right)
  \psi_j \left( \bm{r} \right)
\end{equation}
can also be combined in the following form:
$\rho_{+} \left( \bm{r} \right) = \rho_{\text{V,gs}} \left( \bm{r} \right) + \rho_{\text{S,gs}} \left( \bm{r} \right)$ and
$\rho_{-} \left( \bm{r} \right) = \rho_{\text{V,gs}} \left( \bm{r} \right) - \rho_{\text{S,gs}} \left( \bm{r} \right)$, so that 
\begin{equation}
  \label{eq:psi-dagger}
  \sum_{j \in \text{occ}}
  \psi_j^\dagger \left( \bm{r} \right)
  \left( \bm{\alpha} \cdot \bm{p} - \epsilon_j \right)
  \psi_j \left( \bm{r} \right)
  -
  \frac{1}{2} \left( m - V_{-} \left( \bm{r} \right) \right) \rho_{-} \left( \bm{r} \right)
  +
  \frac{1}{2} \left( m + V_{+} \left( \bm{r} \right) \right) \rho_{+} \left( \bm{r} \right)
  =
  0. 
\end{equation}
If Eq.~(\ref{eq:Dirac2}) is multiplied with
$\bar{\psi}_j = \psi_j^{\dagger} \beta$,
the following expression is obtained:
\begin{equation}
  \label{eq:psi-bar}
  \sum_{j \in \text{occ}}
  \bar{\psi}_j \left( \bm{r} \right)
  \left( \bm{\alpha} \cdot \bm{p} - \epsilon_j \right)
  \psi_j \left( \bm{r} \right)
  +
  \frac{1}{2} \left( m - V_{-} \left( \bm{r} \right) \right) \rho_{-} \left( \bm{r} \right)
  +
  \frac{1}{2} \left( m + V_{+} \left( \bm{r} \right) \right) \rho_{+} \left( \bm{r} \right)
  =
  0. 
\end{equation}
Finally, by adding and subtracting Eqs.~(\ref{eq:psi-dagger}) and (\ref{eq:psi-bar}), we derive
\begin{subequations}
  \begin{align}
    \sum_{j \in \text{occ}}
    \left( \psi_j^\dagger \left( \bm{r} \right) + \bar{\psi}_j \left( \bm{r} \right) \right)
    \left( \bm{\alpha} \cdot \bm{p} - \epsilon_j \right)
    \psi_j \left( \bm{r} \right)
    +
    \left( m + V_{+} \left( \bm{r} \right) \right)
    \rho_{+} \left( \bm{r} \right)
    & =
      0,
      \label{eq:rhop} \\
    \sum_{j \in \text{occ}}
    \left( \psi_j^\dagger \left( \bm{r} \right) - \bar{\psi}_j \left( \bm{r} \right) \right)
    \left( \bm{\alpha} \cdot \bm{p} - \epsilon_j \right)
    \psi_j \left( \bm{r} \right)
    % +
    % \left( V_{-} \left( \bm{r} \right) - m \right)
    -
    \left( m - V_{-} \left( \bm{r} \right) \right)
    \rho_{-} \left( \bm{r} \right)
    & =
      0, 
      \label{eq:rhom}
  \end{align}
\end{subequations}
from which the KS potentials $V_{+}$ and $V_{-}$ are expressed 
\begin{subequations}
  \begin{align}
    V_{+} \left( \bm{r} \right)
    & =
      - m
      -
      \frac{1}{\rho_{+} \left( \bm{r} \right)}
      \sum_{j \in \text{occ}}
      \left( \psi_j^\dagger \left( \bm{r} \right) + \bar{\psi}_j \left( \bm{r} \right) \right)
      \left( \bm{\alpha} \cdot \bm{p} - \epsilon_j \right)
      \psi_j \left( \bm{r} \right),
      \label{eq:Vp} \\
    V_{-} \left( \bm{r} \right)
    & =
      +
      m
      -
      \frac{1}{\rho_{-} \left( \bm{r} \right)}
      \sum_{j \in \text{occ}}
      \left( \psi_j^\dagger \left( \bm{r} \right) - \bar{\psi}_j \left( \bm{r} \right) \right)
      \left( \bm{\alpha} \cdot \bm{p} - \epsilon_j \right)
      \psi_j \left( \bm{r} \right).
      \label{eq:Vm}
  \end{align}
\end{subequations}
The set of IKS equations~(\ref{eq:Vp}) and (\ref{eq:Vm}) can be solved iteratively.
If we assume that the densities in the denominator are the exact (target) densities,
and use Eqs.~(\ref{eq:rhop}) and~(\ref{eq:rhom}) in the numerator to define the densities and potentials for the $n$-th step, the resulting  potentials in the $ \left( n+1 \right)$-th step read
% \begin{equation}
%   \sum_{j \in \text{occ}}{(\psi_j^{(n)\dagger} + \bar{\psi}^{(n)}_j)(\bm{\alpha} \cdot \bm{p}-\epsilon^{(n)}_j)\psi_j^{(n)}}+ V^{(n)}_+ \rho_+ =0,
% \end{equation}
% \begin{equation}
%   \sum_{j \in \text{occ}}{(\psi_j^\dagger - \bar{\psi}_j)(\bm{\alpha} \cdot \bm{p}-\epsilon_j)\psi_j}+ (V_- -2m) \rho_- =0.
% \end{equation}
\begin{align}
  V_{+}^{(n+1)} \left( \bm{r} \right)
  & =
    \frac{\rho_{+}^{(n)} \left( \bm{r} \right)}{\rho_{+} \left( \bm{r} \right)}
    V_{+}^{(n)} \left( \bm{r} \right)
    +
    m \frac{\rho_{+}^{(n)} \left( \bm{r} \right) - \rho_{+} \left( \bm{r} \right)}{\rho_{+} \left( \bm{r} \right)} ,
  \label{eq:Vp-n} \\
  V_{-}^{(n+1)} \left( \bm{r} \right)
  & =
    \frac{\rho_{-}^{(n)} \left( \bm{r} \right)}{\rho_{-} \left( \bm{r} \right)}
    V_{-}^{(n)} \left( \bm{r} \right)
    -
    m \frac{\rho_{-}^{(n)} \left( \bm{r} \right) - \rho_{-} \left( \bm{r} \right)}{\rho_{-} \left( \bm{r} \right)} .
  \label{eq:Vm-n}
\end{align}
\par
In actual IKS calculations, we have modified an algorithm proposed in Ref.~\cite{Jensen.18},
and replaced Eqs.~(\ref{eq:Vp-n}) and (\ref{eq:Vm-n})  with
\begin{align}
  V_{+}^{(n)} \left( \bm{r} \right)
  & =
    V_{+}^{(n-1)} \left( \bm{r} \right)
    +
    \gamma_{+}
    \frac{\rho_{+}^{(n)} \left( \bm{r} \right) - \rho_{+} \left( \bm{r} \right)}{\rho_{+} \left( \bm{r} \right)} ,
  \label{eq:Vp-n-jensen} \\
  V_{-}^{(n)} \left( \bm{r} \right)
  & =
    V_{-}^{(n-1)} \left( \bm{r} \right)
    +
    \gamma_{-}
    \frac{\rho_{-}^{(n)} \left( \bm{r} \right) - \rho_{-} \left( \bm{r} \right)}{\rho_{-} \left( \bm{r} \right)}  .
  \label{eq:Vm-n-jensen}
\end{align}
This algorithm was also used in the first nuclear IKS calculation with non-relativistic EDFs \cite{Accorto.20}, and justified by the following argument.
Equation~(\ref{eq:Vp-n}) for the potential
$V_{+} \left( \bm{r} \right) = V \left( \bm{r} \right) + S \left( \bm{r} \right)$,
which is the equivalent of the non-relativistic KS potential, has a simple interpretation: the potential is enhanced in absolute value 
in those regions where the density is larger than the target density, and reduced in regions where the density is smaller than the target density. 
However, this is what one expects for repulsive potentials (e.g., the Coulomb potential for electrons), whereas in the case of attractive potentials (e.g., the nuclear potential for nucleons)
the opposite should happen.
In Ref.~\cite{Accorto.20}, it has been shown that this issue can be avoided by adopting the modified
algorithm of Eqs.~(\ref{eq:Vp-n-jensen}) and (\ref{eq:Vm-n-jensen}).
Following Ref.~\cite{Accorto.20}, here  we use the value of $1 \, \mathrm{MeV}$
for both parameters $\gamma_{+}$ and $\gamma_{-}$.
In actual calculations we have encountered some stability issues for large values of the
radial coordinate, due to small values of denominators in Eqs.~(\ref{eq:Vp-n-jensen}) and (\ref{eq:Vm-n-jensen}) beyond the nuclear radius.
This problem can be simply solved 
by introducing a cut-off radius $r_{\text{cut}}$,
and setting the potentials to zero for $r>r_{\text{cut}}$.
For the initial KS potential, a realistic Woods-Saxon potential~\cite{Koepf.91} is used,   
and the Broyden mixing procedure~\cite{Baran.08} is employed to solve Eqs.~(\ref{eq:Vp-n-jensen}) and (\ref{eq:Vm-n-jensen}).
The convergence criterion used to halt the iterative IKS algorithm is defined in terms of the absolute variation of the potential, i.e.
\begin{equation}
  \Delta V^{(n)}_{\pm}
  \equiv
  \max_r \left[ V^{(n+1)}_{\pm} \left( r \right) - V^{(n)}_{\pm} \left( r \right) \right]
  <
  \alpha_{\pm}.
\end{equation}
% 
% -----------------------------------------------------------------------------------------------------------------------
\section{A test case}
\label{sec:test}
% -----------------------------------------------------------------------------------------------------------------------
\par
The atomic nucleus is a complex quantum mechanical system with two types of constituent particles of spin one-half and, therefore, a general EDF will be a functional of isoscalar, isovector, and spin densities, as well as corresponding currents.
Even in the simplest case of spin-saturated even-even nuclei, the EDF will depend on isoscalar and isovector densities.
The problem, of course, is that accurate data exists only on charge (proton) densities and, thus, adjusting a general functional to empirical densities is not a straightforward procedure.
In the particular case of relativistic EDFs that we consider here, the functional depends also on the Lorentz scalar single-nucleon density, which is not an observable.
In the final section, we will discuss a possible approach that can be used to determine the scalar and isovector densities in an indirect way,
but, for the purpose of testing the proposed method, here only $N=Z$ systems without Coulomb interaction are considered.
For such artificial nuclei, to demonstrate the relativistic IKS method with DFPT, 
we will use an existing relativistic EDF as the \textit{exact} target functional, and apply the method developed in the previous section to improve different approximate functionals towards the target functional.
In real nuclei, the exact functional is, of course, unknown and we will need more than ground-state data to determine the functional dependence on various nuclear densities. 
\par
For the \textit{exact} target EDF, we will use the relativistic EDF DD-PC1 \cite{Nik.08}, for which the single-nucleon Hamiltonian reads
\begin{equation}
  \label{eq:hamiltonian}
  \hat{h}
  =
  \bm{\alpha} \cdot \bm{p}
  +
  \beta \left( m + S \left( \bm{r} \right) \right)
  +
  V_0 \left( \bm{r} \right)
  +
  \Sigma_{\text{R}} \left( \bm{r} \right),
\end{equation}
where
the scalar potential, vector potential, and rearrangement terms are respectively defined by 
\begin{align}
  S
  & =
    \alpha_{\text{S}} \left( \rho \right) \rho_{\text{S}}
    +
    \delta_{\text{S}} \triangle \rho_{\text{S}},
    \notag \\
  V
  & =
    \alpha_{\text{V}} \left( \rho \right) \rho_{\text{V}}
    +
    \alpha_{\text{TV}} \left( \rho \right) \vec{\rho}_{\text{TV}} \cdot \vec{\tau} 
    +
    e
    \frac{1 - \tau_3}{2} A_{0},
    \notag \\
  \Sigma_{\text{R}}
  & =
    \frac{1}{2}
    \frac{\partial \alpha_{\text{S}}}{\partial \rho} \rho_{\text{S}}^2
    +
    \frac{1}{2}
    \frac{\partial \alpha_{\text{V}}}{\partial \rho} \rho_{\text{V}}^2
    +
    \frac{1}{2}
    \frac{\partial \alpha_{\text{TV}}}{\partial \rho} \rho_{\text{TV}}^2.
\end{align}
In these expressions, 
$m$ is the nucleon mass,
$\alpha_{\text{S}} \left( \rho \right)$,
$\alpha_{\text{V}} \left( \rho \right)$, and
$\alpha_{\text{TV}} \left( \rho \right)$ are
density-dependent couplings for different space-isospace channels, 
$\delta_{\text{S}}$ is the coupling constant of the derivative term,
$e$ is the electric charge,
$\frac{1 - \tau_3}{2} A_{0}$ corresponds to the Coulomb interaction,
and the single-nucleon densities 
$\rho_{\text{S}}$ (scalar-isoscalar density),
$\rho_{\text{V}}$ (time-like component of the isoscalar current),
and $\rho_{\text{TV}}$  (time-like component of the isovector current).
\par
In addition to contributions from the isoscalar-vector four-fermion
interaction and the electromagnetic interaction, the isoscalar-vector
self-energy includes the \textit{rearrangement} terms in 
$\Sigma_{\text{R}}$ that arise from the variation of the vertex functionals
$\alpha_{\text{S}}$, $\alpha_{\text{V}}$, and $\alpha_{\text{TV}}$ with respect to the nucleon
fields in the vector density operator $\hat{\rho}_{\text{V}}$.
\par
In a phenomenological construction of a relativistic EDF,
one starts from an assumed \textit{ansatz} for the medium dependence
of the mean-field nucleon self-energies, and adjusts the free
parameters directly to ground-state data of finite nuclei.
Guided by
the microscopic density dependence of the vector and scalar
self-energies, the following practical \textit{ansatz} for the functional form
of the couplings was adopted in Ref.~\cite{Nik.08}:
\begin{align}
  \alpha_{\text{S}} \left( \rho_\text{V} \right)
  & =
    a_{\text{s}} + \left( b_{\text{s}} + c_{\text{s}} x \right) e^{-d_{\text{s}} x},
    \notag \\
  \alpha_{\text{V}} \left( \rho_\text{V} \right)
  & = 
    a_{\text{v}} + b_{\text{v}} e^{-d_{\text{v}} x},
    \label{parameters} \\
  \alpha_{\text{TV}} \left( \rho_\text{V} \right)
  & =
    b_{\text{tv}} e^{-d_{\text{tv}} x},
    \notag 
\end{align}
with $x=\rho_\text{V}/\rho_{\text{sat}}$, where $\rho_{\text{sat}}$ denotes the nucleon
density at saturation in symmetric nuclear matter.
The set of 10 parameters was adjusted in a $\chi^2$ fit
to the experimental masses of 64 axially deformed nuclei
in the mass regions $A\approx 150 $--$180$ and $A \approx 230$--$250$.
The resulting functional DD-PC1 \cite{Nik.08} has been further tested in
calculations of binding energies, charge radii, deformation
parameters, neutron-skin thickness, and excitation energies of giant
monopole and dipole resonances.
During the last decade the functional DD-PC1 has 
successfully been applied in a number of studies of various nuclear phenomena, from ground-state properties to the description of collective spectra, giant resonances, shape-phase transitions, and the dynamics of nuclear fission. 
\par
In the simplified case of $N=Z$ doubly closed-shell nuclei without Coulomb interaction, there is no contribution of the isovector channel either.
We will use
four $N=Z$ systems: $^{8}_{8}16$, $^{20}_{20}40$, $^{28}_{28}56$, and $^{50}_{50}100$ to improve, 
starting from the \textit{exact} ground-state densities,
the approximate zeroth-order functionals towards DD-PC1.
First, we illustrate the accuracy of the IKS scheme, described in the previous section, in determining the KS potentials for given scalar and vector densities of the $N=Z=8$ system. 
The densities-to-potentials inversion enables the calculation of the single-particle energies that appear on the right-hand side Eq.~(\ref{eq:final}). 
\par
\begin{figure}[htb]
  \centering
  \includegraphics[scale=0.65]{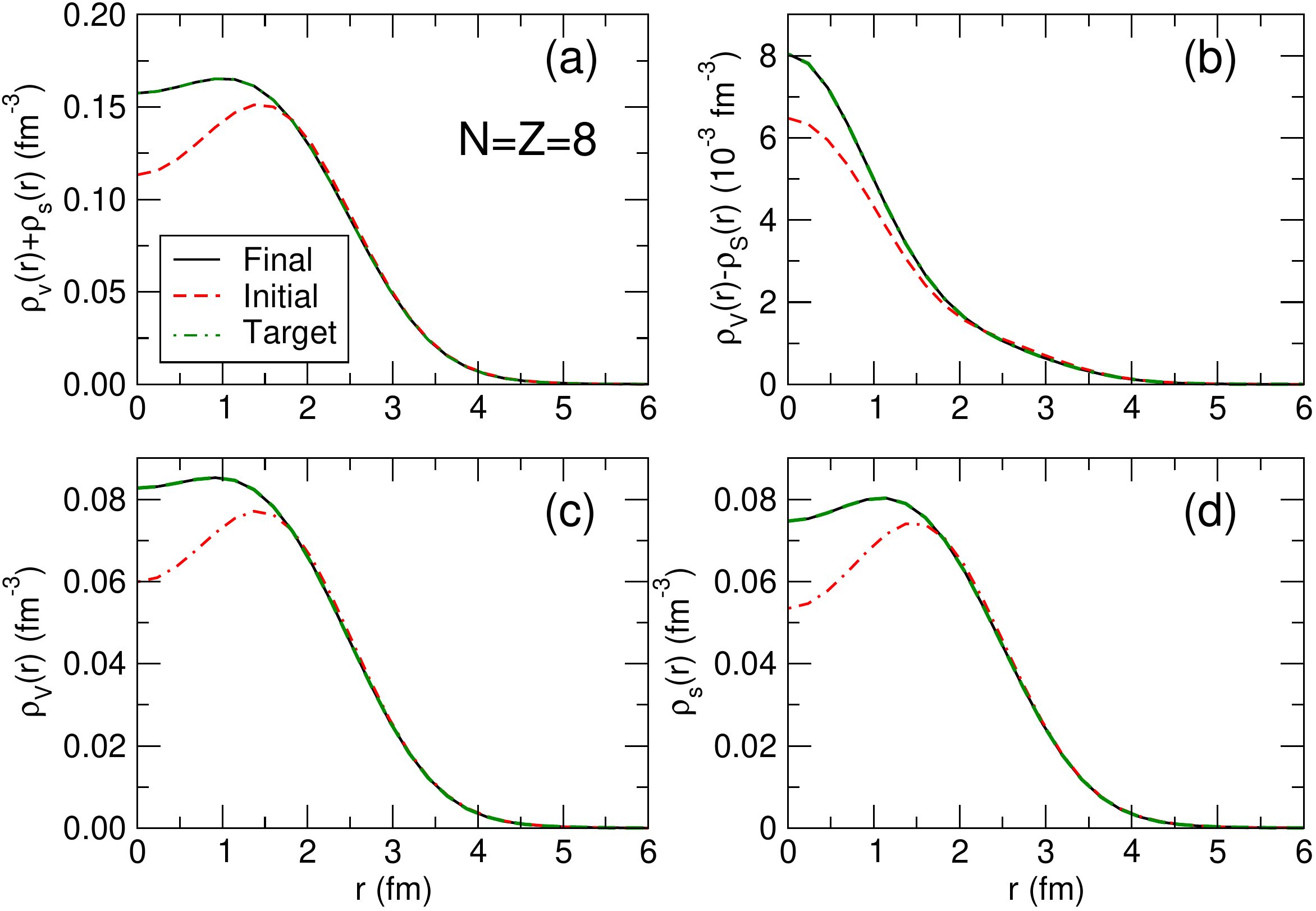}
  \caption{
    (Color online)
    (a) The sum of neutron vector and scalar densities in the $N=Z=8$ system as a function of the radial coordinate.
    The target density obtained using the DD-PC1 functional (dashed green curve) is compared to the 
    the density calculated in the initial step of the inversion method (dot-dashed red) with a Woods-Saxon potential, and to the final IKS density (solid black).
    (b) Same as in panel (a) but for the difference between the neutron vector and scalar densities.
    (c) Same as in panel (a) but for the neutron vector density.
    (d) Same as in panel (a) but for the neutron scalar density.}
  \label{fig_IKS_dens_o16}
\end{figure}
\begin{figure}[htb]
  \centering
  \includegraphics[scale=0.65]{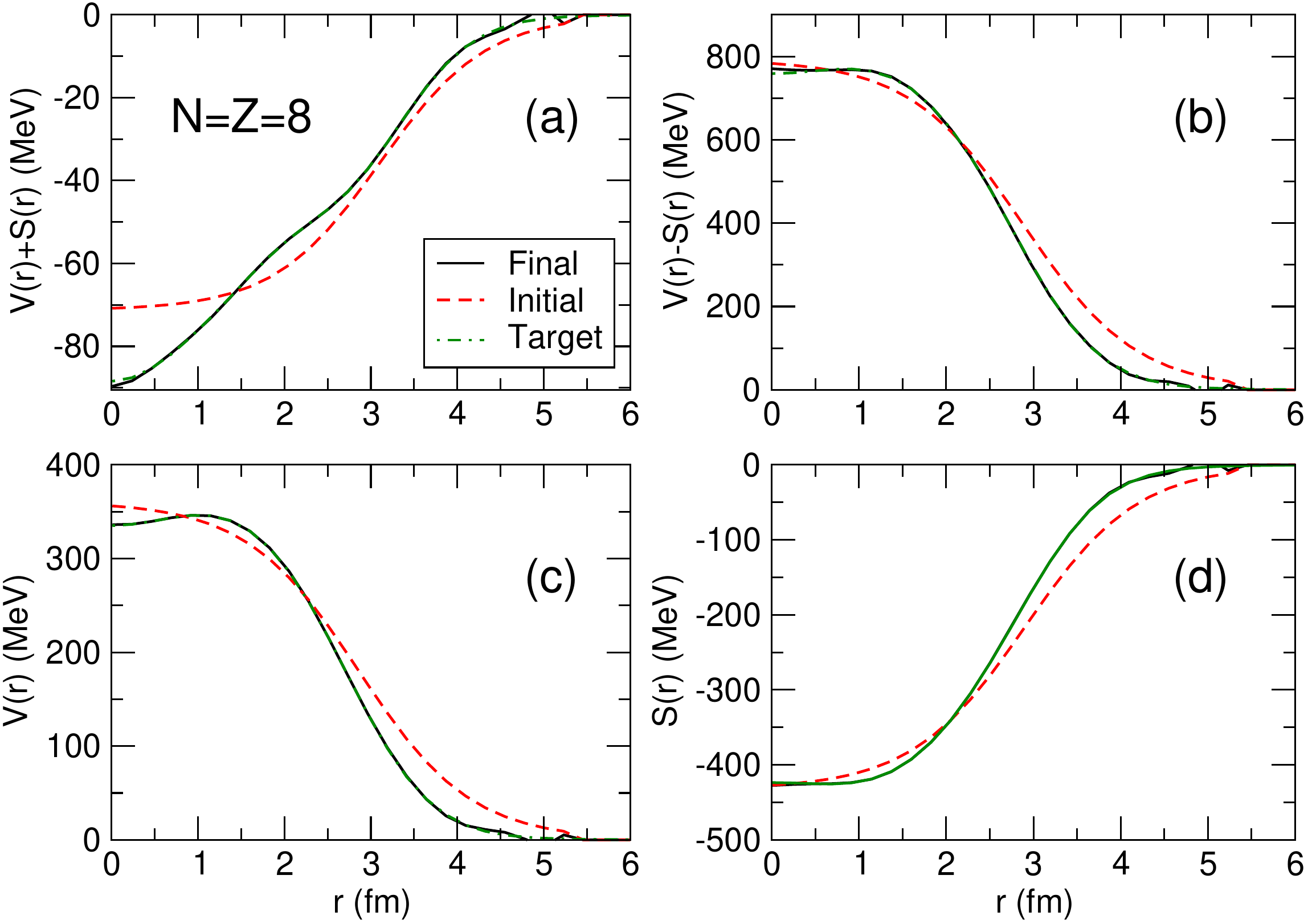}
  \caption{
    (Color online)
    (a) The sum of neutron vector and scalar potentials in the $N=Z=8$ system as a function of the radial coordinate.
    The target DD-PC1 Kohn-Sham potential (dashed green curve) is compared to the 
    initial Woods-Saxon potential (dot-dashed red), and to the final IKS potential (solid black).
    (b) Same as in panel (a) but for the difference between the neutron vector and scalar potentials.
    (c) Same as in panel (a) but for the neutron vector potential.
    (d) Same as in panel (a) but for the neutron scalar potential.}
  \label{fig_IKS_pot_o16}   
\end{figure}

Figure \ref{fig_IKS_dens_o16} compares the densities obtained in the inverse KS scheme to the target DD-PC1 densities.
We plot four different neutron densities: the sum of the scalar and vector density
$\rho_{+} \left( r \right) = \rho_{\text{V}} \left( r \right) + \rho_{\text{S}} \left( r \right)$
(panel (a)),
the difference between the vector and scalar density
$\rho_{-} \left( r \right) = \rho_{\text{V}} \left( r \right) - \rho_{\text{S}} \left( r \right)$
(panel (b)),
and separately the vector $\rho_{\text{V}} \left( r \right)$ (panel (c))
and scalar $\rho_{\text{S}} \left( r \right)$ densities (panel (d)).
Without Coulomb interaction the proton densities are, of course, identical to the neutron ones.
In all four panels the dash-dotted green curves denote the target densities calculated with the DD-PC1 functional, the dashed red curves are the initial densities that correspond to Woods-Saxon potentials and, finally, the solid black curves represent the final densities obtained by the inversion method. 
The corresponding results for the potentials are shown in Fig.~\ref{fig_IKS_pot_o16}:
the sum of the vector and scalar potentials
$ V_{+} \left( r \right) = V \left( r \right) + S \left( r \right) $ (panel (a)),
the difference between the vector and scalar potential
$ V_{-} \left( r \right) = V \left( r \right) - S \left( r \right) $ (panel (b)),
and separately the vector $ V \left( r \right) $ (panel (c))
and scalar $ S \left( r \right) $ (panel (d)) potentials.
Again, the green dash-dotted curves denote the target potentials, the dashed red curves are the initial Woods-Saxon potentials, and
the solid black curves are the final potentials obtained by the inversion method.
Obviously, the result is that one cannot distinguish between the target and final IKS densities and potentials.
The latter can, therefore, be used to calculate the single-particle energies that are needed in the application of the DFPT method. 
%%% 
%%% 
%%% 
% ---------------------------------------------------------------------------------------------------------------------------
\par
In the test case, we will assume for the known functional
$E_{\text{int}}^{(0)}\left[ \rho_{\text{V}}, \rho_{\text{S}} \right]$
a simple form that is actually a part of the DD-PC1 functional
\begin{equation}
  \label{E_0}
  E^{(0)}_{\text{int}} \left[ \rho_{\text{V}},\rho_{\text{S}} \right]
  =
  \frac{1}{2} \alpha_{\text{s}}^{(0)} \rho_{\text{S}}^2
  +
  \frac{1}{2} \alpha_{\text{v}}^{(0)} \rho_{\text{V}}^2
  +
  \delta_{\text{s}} \rho_{\text{s}} \triangle \rho_{\text{s}}, 
\end{equation}
The values for the $a_{\text{s}}^{(0)}$ and $a_{\text{v}}^{(0)}$ constants are those used in the DD-PC1 functional:
$a_{\text{s}}^{(0)}=a_{\text{s}} =-10.4602 \, \mathrm{fm}^{-2}$ and
$a_{\text{v}}^{(0)}=a_{\text{v}} =  5.9195 \, \mathrm{fm}^{-2}$,
and the same choice is made for the derivative term: $\delta_{\text{s}}=-0.8149$.
Note that the first two terms of this functional coincide with the simple Walecka mean-field model which, with only two parameters, produces a realistic equation of state of symmetric nuclear matter. Such a model, in fact, correspond to a local density approximation (LDA) for the EDF.
The derivative term is used in modeling finite systems and takes into account the rapid variations of the density in the surface region.
The strength parameter of this term can be determined, at least qualitatively, from microscopic calculations of inhomogeneous nuclear matter. 
\par
For the remaining \textit{unknown} part of the functional $E_{\text{int}}^{(1)}\left[ \rho_{\text{V}}, \rho_{\text{S}} \right]$ we choose 
\begin{equation}
  \label{E_1}
  E^{(1)}_{\text{int}} \left[ \rho_{\text{V}},\rho_{\text{S}} \right]
  =
  \frac{1}{2}
  \alpha_{\text{s}}^{(1)} \left( \rho_{\text{V}} \right)
  \rho_{\text{S}}^2
  +
  \frac{1}{2}
  \alpha_{\text{v}}^{(1)} \left( \rho_{\text{V}} \right)
  \rho_{\text{V}}^2.
\end{equation}
where
$\alpha_{\text{s}}^{(1)} \left( \rho_{\text{V}} \right) $ and
$\alpha_{\text{v}}^{(1)} \left( \rho_{\text{V}} \right) $ have the functional form of the density-dependent parts of the DD-PC1 couplings:
\begin{equation}
  \alpha_{\text{s}}^{(1)} \left( \rho_{\text{V}} \right)
  =
  \left(
    b_{\text{s}}^{(1)} + c_{\text{s}}^{(1)} x
  \right)
  e^{-d_{\text{s}} x}
  \quad \text{and} \quad
  \alpha_{\text{v}}^{(1)} \left( \rho_{\text{V}} \right)
  =
  b_{\text{v}}^{(1)}
  e^{-d_{\text{v}} x},
\end{equation}
with $x=\rho_{\text{V}}/\rho_{\text{sat}}$, and $\rho_{\text{sat}}=0.152 \, \mathrm{fm}^{-3}$.
As explained above, the parameters of the functional DD-PC1 were adjusted to reproduce the nuclear matter equation of state and the experimental masses of 64 deformed nuclei.
The test of the method proposed in this work consist in trying to determine the parameters of
$E_{\text{int}}^{(1)}\left[ \rho_{\text{V}}, \rho_{\text{S}} \right]$ shown in Eq.~(\ref{E_1})
(i.e., $ b_{\text{s}}^{(1)} $, $ c_{\text{s}}^{(1)} $, and $ b_{\text{v}}^{(1)} $) 
by using density functional perturbation theory and the IKS scheme, that is, using Eq.~(\ref{eq:final}).
Because the right-hand side of this equation is just a number that can be evaluated provided the exact single-particle energies and vector and scalar densities are known, a different finite system is needed for each parameter of the unknown functional.
Since this is an illustrative test, we will employ three
$N=Z$ systems: $^{8}_816$, $^{28}_{28}56$ and $^{50}_{50}100$ to determine the constants 
$b_{\text{s}}^{(1)}$, $c_{\text{s}}^{(1)}$ and $b_{\text{v}}^{(1)}$,
while $d_{\text{s}}$ and $d_{\text{v}}$ are again fixed to the DD-PC1 values.
Note that, even though the problem has been simplified to a certain extent, nevertheless the test is far from being trivial.
Namely, only three artificial systems are used to reproduce the values of parameters that were originally adjusted to the experimental masses of a large number of nuclei.
Moreover, since our choice for the unperturbed functional is obviously not close to the
exact target functional, it is far from obvious that a first-order perturbation method will determine the unknown parameters with sufficient accuracy.
Hence, we repeat the calculation in several iterative steps, as described in the previous section [Eq.~(\ref{eq:iter})],
and at each step improve the values of $b_{\text{s}}^{(1)}$, $c_{\text{s}}^{(1)}$, and $b_{\text{v}}^{(1)}$.
Note that, because of the functional form of $ E_{\text{int}}^{(1)} $ expressed in Eq.~(\ref{E_1}), 
\begin{align}
  E_{\text{int}}^{\text{$(1)$, $ n $-th}} \left[ \rho_{\text{V}}, \rho_{\text{S}} \right]
  & =
    \sum_{k=1}^{n}
    \left[
    \frac{1}{2}
    \left(
    b_{\text{s}}^{\text{$(1)$, $ k $-th}} + c_{\text{s}}^{\text{$(1)$, $ k $-th}} x
    \right)
    e^{-d_{\text{s}} x}
    \rho_{\text{S}}^2
    +
    \frac{1}{2}
    b_{\text{v}}^{\text{$(1)$, $ k $-th}}
    \rho_{\text{V}}^2
    \right]
    \notag \\
  & =
    \frac{1}{2}
    \left(
    \left\{
    \sum_{k=1}^{n}
    b_{\text{s}}^{\text{$(1)$, $ k $-th}}
    \right\}
    +
    \left\{
    \sum_{k=1}^{n}
    c_{\text{s}}^{\text{$(1)$, $ k $-th}}
    \right\}
    x
    \right)
    e^{-d_{\text{s}} x}
    \rho_{\text{S}}^2
    +
    \frac{1}{2}
    \left\{
    \sum_{k=1}^{n}
    b_{\text{v}}^{\text{$(1)$, $ k $-th}}
    \right\}
    \rho_{\text{V}}^2
\end{align}
holds.
Hence, hereafter,
$ \left( b_{\text{s}}^{(1)} \right)_i $,
$ \left( c_{\text{s}}^{(1)} \right)_i $, and
$ \left( b_{\text{v}}^{(1)} \right)_i $
simply denote
$ \left\{ \sum_{k=1}^{i} b_{\text{s}}^{\text{$(1)$, $ k $-th}} \right\} $,
$ \left\{ \sum_{k=1}^{i} c_{\text{s}}^{\text{$(1)$, $ k $-th}} \right\} $, and
$ \left\{ \sum_{k=1}^{i} b_{\text{v}}^{\text{$(1)$, $ k $-th}} \right\} $,
respectively. 
\par
\begin{figure*}[!ht]
  \includegraphics[scale=0.5]{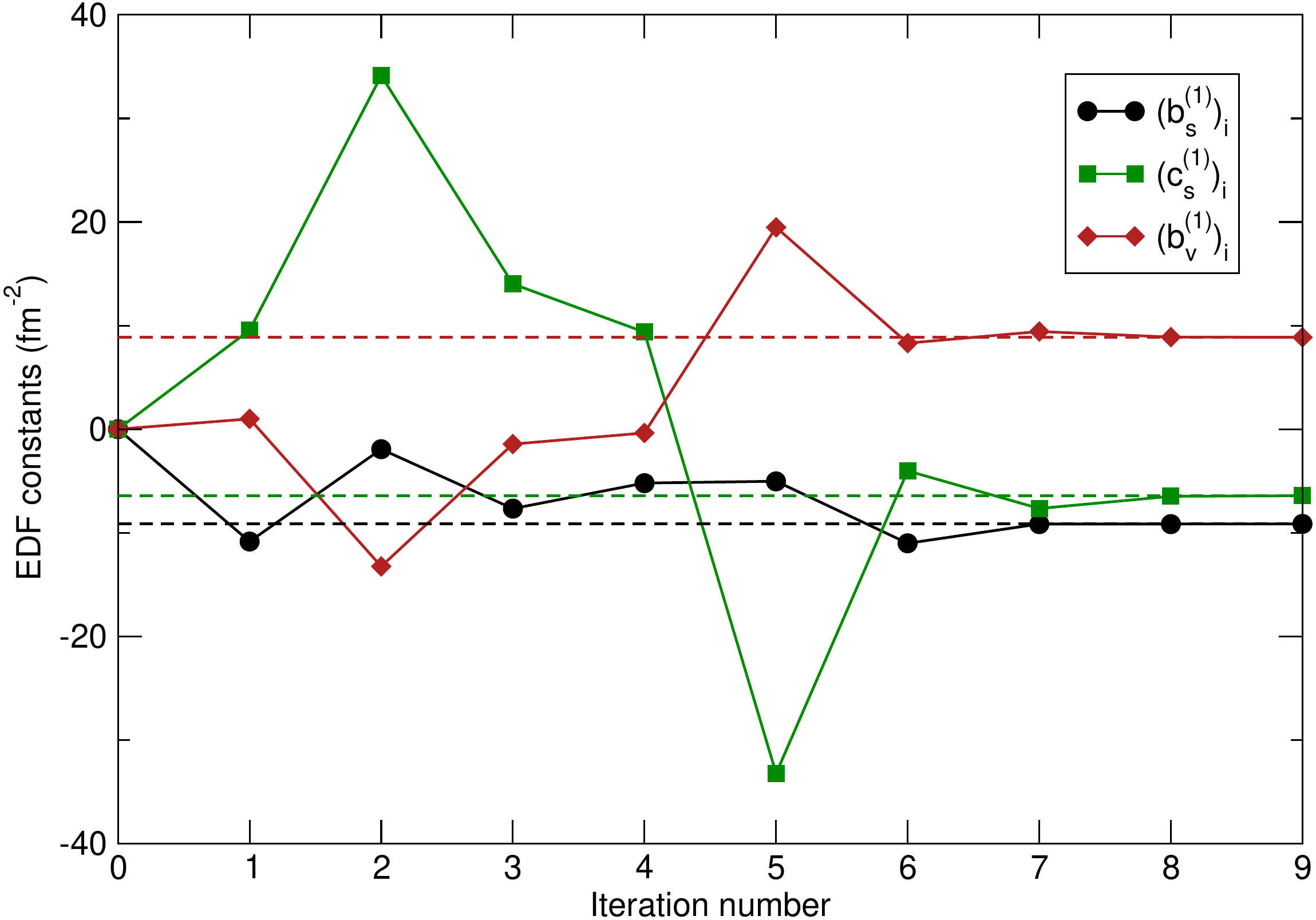}
  \caption{
    (Color online).
    Values of the constants
    $ \left( b_{\text{s}}^{(1)} \right)_i $,
    $ \left( c_{\text{s}}^{(1)} \right)_i $ and
    $ \left( b_{\text{v}}^{(1)} \right)_i $ 
    at different iteration steps. The dashed lines denote the target values that correspond to the functional DD-PC1.}
  \label{fig:constants_dfpt_test}
\end{figure*}

Figure \ref{fig:constants_dfpt_test} displays the values of
$ \left( b_{\text{s}}^{(1)} \right)_i $,
$ \left( c_{\text{s}}^{(1)} \right)_i $, and
$ \left( b_{\text{v}}^{(1)} \right)_i $ at each iteration step $i$.
Assuming that nothing is known about these parameters, we start with zero values.
After some initial oscillations in the first few steps, especially between 
$c_{\text{s}}^{(1)}$ and $b_{\text{v}}^{(1)}$, the parameters converge to the values that correspond to the DD-PC1 target functional, denoted by the horizontal lines in Fig.~\ref{fig:constants_dfpt_test}.
The results of this test demonstrate not only the feasibility of the $ \text{IKS} + \text{DFPT} $ method for nuclear densities, but also the convergence and accuracy of the iteration scheme. 
%%% 
% 
% ----------------------------------------------------------
\section{An illustrative calculation}
\label{sec:example}
% ----------------------------------------------------------
\par
In the second example,
we again use DD-PC1 as the unknown target functional, and the corresponding exact single-particle energies that appear on the right-hand side of Eq.~(\ref{eq:final}) are obtained by the IKS method as described in Sec.~\ref{subsec:IKS}.
Also for the known functional
$E_{\text{int}}^{(0)} \left[ \rho_{\text{V}}, \rho_{\text{S}} \right]$
the simple form of Eq.~(\ref{E_0}) is adopted, that is 
\begin{equation}
  E^{(0)}_{\text{int}} \left[ \rho_{\text{V}},\rho_{\text{S}} \right]
  =
  \frac{1}{2} a_{\text{s}} \rho_{\text{S}}^2
  +
  \frac{1}{2} a_{\text{v}} \rho_{\text{V}}^2
  +
  \delta_{\text{s}} \rho_{\text{s}} \triangle \rho_{\text{s}}, 
  \label{E_ini}
\end{equation}
with the DD-PC1 values of the three parameters.
For the remaining unknown part of the functional 
\begin{displaymath}
  E^{(1)}_{\text{int}} \left[ \rho_{\text{V}},\rho_{\text{S}} \right]
  =
  \frac{1}{2} \alpha_{\text{s}}^{(1)} \left( \rho_{\text{V}} \right) \rho_{\text{S}}^2
  +
  \frac{1}{2} \alpha_{\text{v}}^{(1)} \left( \rho_{\text{V}} \right) \rho_{\text{V}}^2, 
  \label{E_int}
\end{displaymath}
we choose a polynomial form of the couplings
$\alpha_{\text{s}}^{(1)} \left( \rho_{\text{V}} \right)$
and
$\alpha_{\text{v}}^{(1)} \left( \rho_{\text{V}} \right)$:
\begin{equation}
  \alpha_{\text{s}}^{(1)} \left(\rho_{\text{V}} \right)
  =
  b_{\text{s}}^{(1)} \left( x-1 \right)
  +
  c_{\text{s}}^{(1)} \left( x-1 \right)^2
  \quad \text{and} \quad
  \alpha_{\text{v}}^{(1)} \left( \rho_{\text{V}} \right)
  =
  b_{\text{v}}^{(1)} \left( x-1 \right)
  +
  c_{\text{v}}^{(1)} \left( x-1 \right)^2 , 
  \label{quad}
\end{equation}
with $x=\rho_{\text{V}}/\rho_{\text{sat}}$,
and $\rho_{\text{sat}}=0.152 \, \mathrm{fm}^{-3}$.
Therefore, we will examine whether the known functional
$E_{\text{int}}^{(0)}\left[ \rho \right]$
can be improved towards the exact target functional DD-PC1, by assuming a polynomial density dependence of the coupling parameters of
$E^{(1)}_{\text{int}}\left[ \rho \right]$.
\par
\begin{figure*}[!ht]
  \includegraphics[scale=0.5]{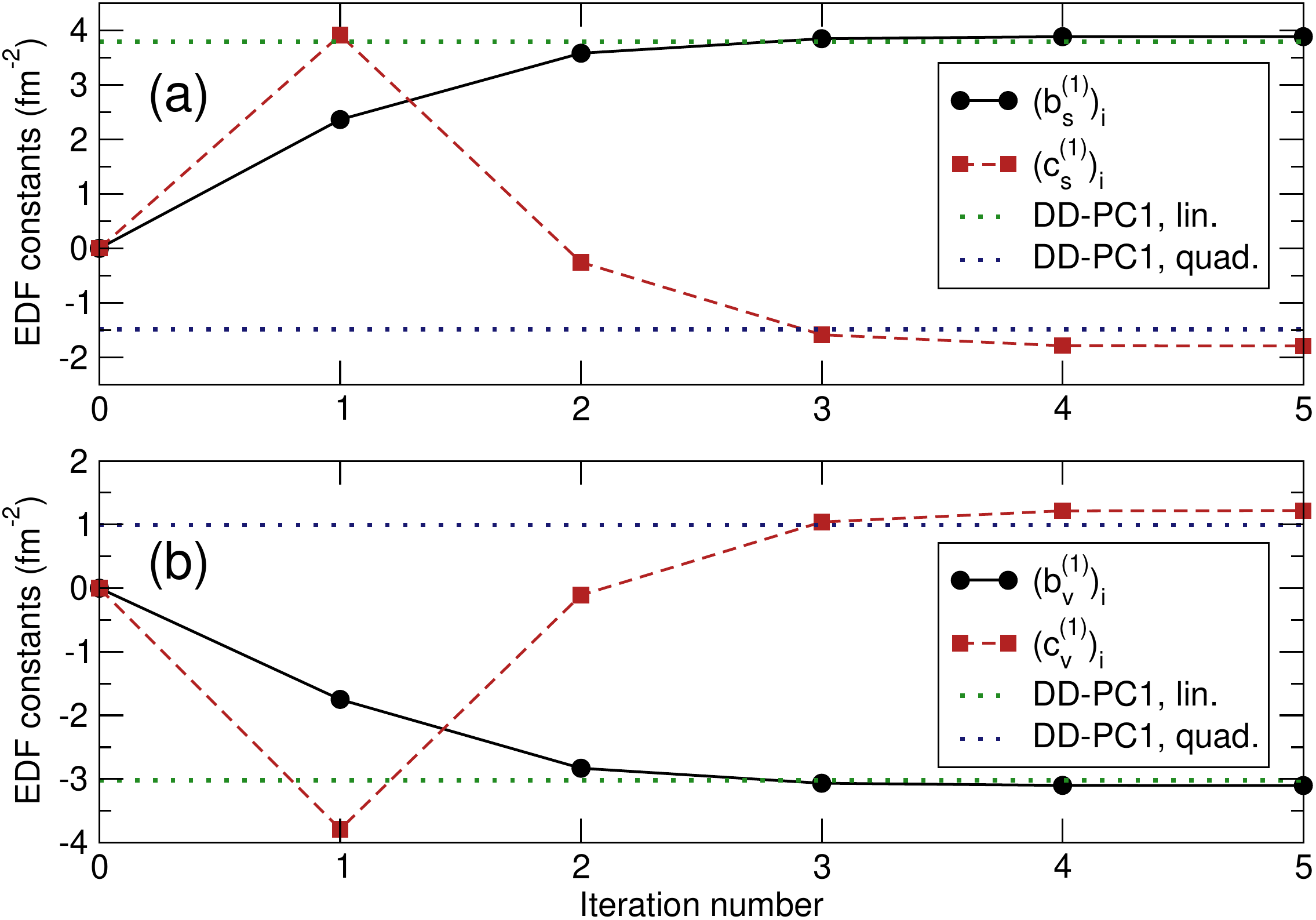}
  \caption{
    (Color online).
    Values of the constants
    $ \left( b_{\text{s}}^{(1)} \right)_i $,
    $ \left( c_{\text{s}}^{(1)} \right)_i $ (panel (a)),
    $ \left( b_{\text{v}}^{(1)} \right)_i $ and
    $ \left( c_{\text{v}}^{(1)} \right)_i $ (panel (b))
    at different iteration steps of the $ \text{IKS} + \text{DFPT} $ calculation.
    The dashed lines denote the corresponding parameters of the linear and quadratic term in the Taylor expansion of the DD-PC1 couplings.}
  \label{fig:4par}
\end{figure*}

Since the values of four parameters have to be determined, Eq.~(\ref{eq:final}) requires the input from four finite systems.
Here we choose: $N=Z=8$, $N=Z=20$, $N=Z=28$, and $N=Z=50$.
In Fig.~\ref{fig:4par} the parameters
$ \left( b_{\text{s}}^{(1)} \right)_i $,
$ \left( c_{\text{s}}^{(1)} \right)_i $,
$ \left( b_{\text{v}}^{(1)} \right)_i $, and
$ \left( c_{\text{v}}^{(1)} \right)_i $ are shown at each iteration step of the $ \text{IKS} + \text{DFPT} $ procedure.
They are compared with the parameters of the linear and quadratic term in the Taylor expansion of the DD-PC1 couplings:
% 
%\begin{equation}
%  \rho_{\text{sat}}
 % \left.
  %  \frac{d \alpha_{\text{s,v}}^{\text{DD-PC1}}}{d\rho_{\text{V}}}
 % \right|_{\rho_{\text{V}}=
 % \rho_{\text{sat}}}
 % \quad \text{and} \quad
 % \frac{1}{2}
 % \rho_{\text{sat}}^2
 % \left.
 %   \frac{d^2 \alpha_{\text{s,v}}^{\text{DD-PC1}}}{d\rho_{\text{V}}^2}
 % \right|_{\rho_{\text{V}} =
 % \rho_{\text{sat}}}.
%\end{equation}
\begin{equation}
\alpha_\text{s,v}^{\text{DD-PC1}}(\rho_\text{V}) \approx \alpha_\text{s,v}(\rho_\text{sat}) +\left.
    \frac{d \alpha_{\text{s,v}}^{\text{DD-PC1}}}{d\rho_{\text{V}}}\right|_{\rho_{\text{V}}= \rho_{\text{sat}}} (\rho_\text{V}-\rho_\text{sat})
    +\frac{1}{2}  \left.   \frac{d^2 \alpha_{\text{s,v}}^{\text{DD-PC1}}}{d\rho_{\text{V}}^2} \right|_{\rho_{\text{V}} = \rho_{\text{sat}}} (\rho_\text{V}-\rho_\text{sat})^2, 
\end{equation}
or expressed in terms of $x=\rho_\text{V}/\rho_{\text{sat}}$: 
\begin{equation}
\alpha_\text{s,v}^{\text{DD-PC1}}(\rho_\text{V}) \approx \alpha_\text{s,v}(\rho_\text{sat}) +\rho_\text{sat}\left.
    \frac{d \alpha_{\text{s,v}}^{\text{DD-PC1}}}{d\rho_{\text{V}}}\right|_{\rho_{\text{V}}= \rho_{\text{sat}}} (x-1)
    +\frac{1}{2} \rho_{\text{sat}}^2 \left.   \frac{d^2 \alpha_{\text{s,v}}^{\text{DD-PC1}}}{d\rho_{\text{V}}^2} \right|_{\rho_{\text{V}} = \rho_{\text{sat}}} (x-1)^2.
\end{equation}

\begin{figure*}[!ht]
  \includegraphics[scale=0.5]{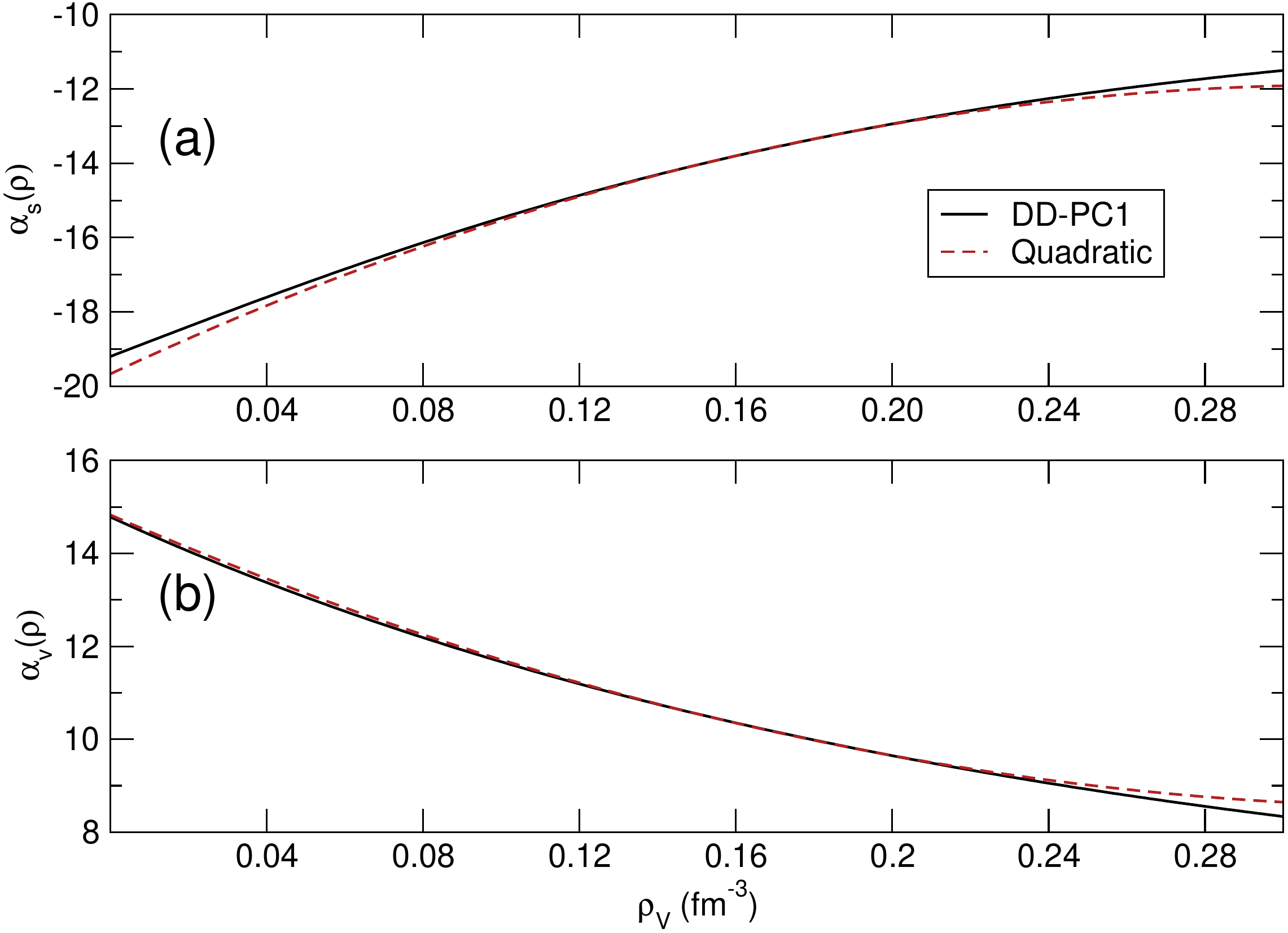}
  \caption{
    (Color online) Scalar (panel (a)) and vector (panel (b)) $ \text{IKS} + \text{DFPT} $ couplings as functions of vector density, compared to the corresponding DD-PC1 target coupling functions.}
  \label{fig:couplings_4par}
\end{figure*}

Even though we start with zero initial values, after only a few iterations the parameters of the linear and quadratic couplings of the unknown functional reach values that are very close to the corresponding parameters of the Taylor expansion of the \textit{target} DD-PC1 couplings. 
%%% 
%%% 
Using the final values of the parameters, we calculate the total scalar and vector couplings 
$\alpha_{\text{S}} \left( \rho_{\text{V}} \right)$ and
$\alpha_{\text{V}} \left( \rho_{\text{V}} \right)$ as functions of the vector density, and compare them with the corresponding couplings of the functional DD-PC1 in Fig.~\ref{fig:couplings_4par}.
While the couplings of the unknown functional $E^{(1)}_{\text{int}}$ have been approximated by simple quadratic functions of the vector density, nevertheless the final $ \text{IKS} + \text{DFPT} $ scalar and vector couplings accurately reproduce the DD-PC1 target couplings over a broad range of densities. 
%%% 
%%% 
\par
\begin{figure*}[!ht]
  \includegraphics[scale=0.5]{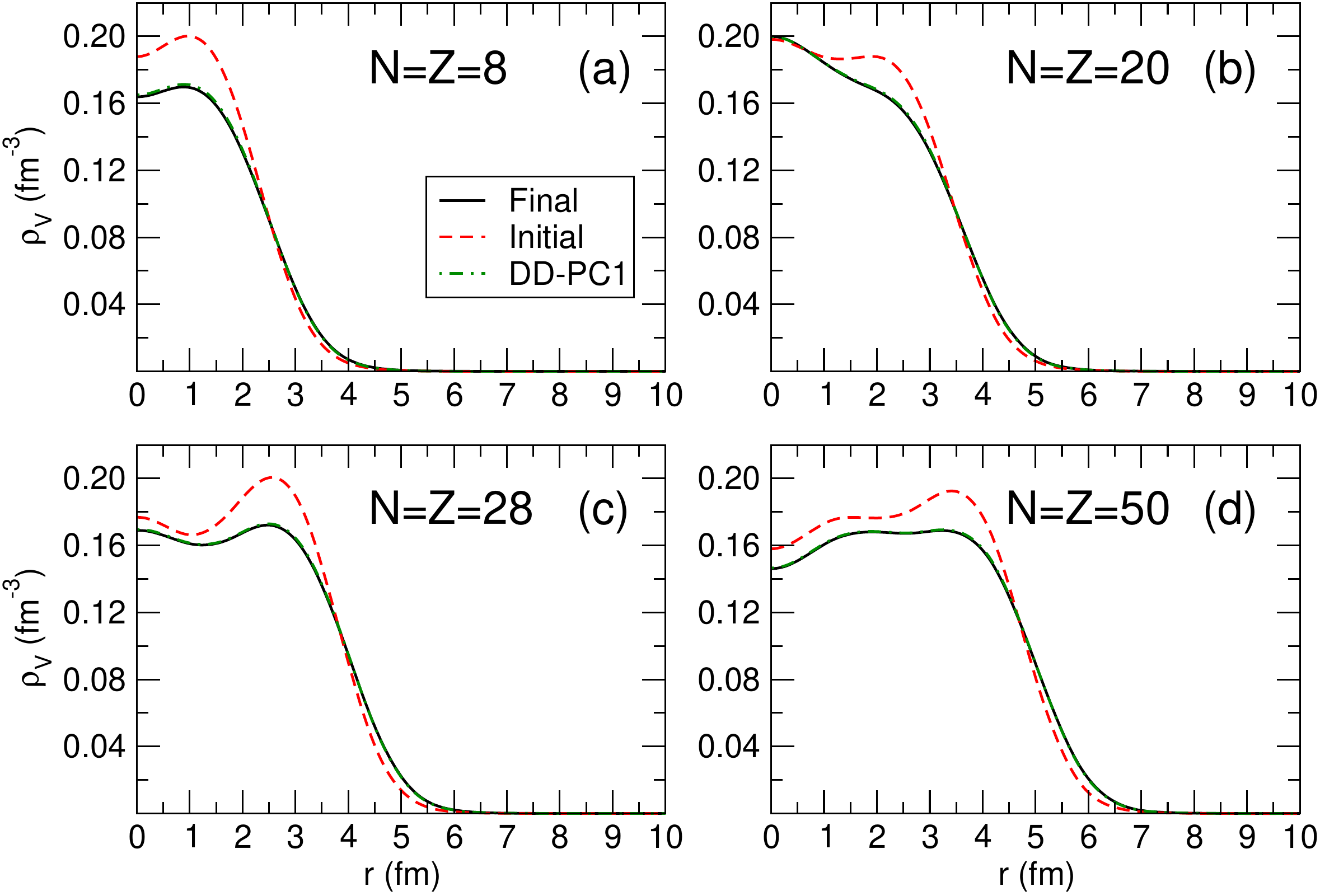}
  \caption{
    (Color online). The vector densities of the four symmetric systems: $N=Z=8$, $N=Z=20$, $N=Z=28$, and $N=Z=50$.
    The dashed red curves are the densities that correspond to the unperturbed initial functional $E_{\text{int}}^{(0)}$ shown in Eq.~(\ref{E_ini}).
    The dot-dashed green and solid black curves denote the densities obtained with the target functional DD-PC1 and the final results of the $ \text{IKS} + \text{DFPT} $ calculation, respectively.}
  \label{fig3}
\end{figure*}

Finally, in Fig.~\ref{fig3} we compare the vector densities for the four symmetric systems:
$N=Z=8$, $N=Z=20$, $N=Z=28$, and $N=Z=50$, calculated with the $ \text{IKS} + \text{DFPT} $ method and the target functional DD-PC1.
The red curves denote the densities that correspond to the unperturbed initial functional $E_{\text{int}}^{(0)}$ and they are, of course, very different from those obtained with the target functional.
However, even when the \textit{unknown} part of the functional is approximated by the simple expressions of Eq.~(\ref{quad}), the $ \text{IKS} + \text{DFPT} $ method produces ground-state densities that are virtually identical to the \textit{exact} target densities.
%%%
%%% 
\section{Summary}
\label{sec:conclusion}
In this study we have considered an interesting problem in the framework of nuclear energy density functionals, namely, how to improve a given functional towards an exact but unknown Kohn-Sham exchange-correlation functional.
Based on the density functional perturbation theory and inverse Kohn-Sham method, a model has been developed that can be used to improve an approximate relativistic EDF. 
\par
Using the method introduced in Ref.~\cite{Naito.19} for non-relativistic functionals, and based on the density functional perturbation theory, we have derived Eq.~(\ref{eq:final}) that is used to compute the first-order correction to an approximate zeroth-order functional.
The input to this equation are the exact ground-state densities and the known functional, as well as the exact single-particle energies that are also implicit functionals of the densities.
We then use the inverse Kohn-Sham (IKS) method to calculate these single-particle energies starting from given ground-state densities. 
\par
In practice, one must assume a certain \textit{ansatz} for the first-order correction to the Kohn-Sham exchange-correlation functional, and use empirical exact ground-state densities of finite systems to determine the corresponding phenomenological parameters.
In case the first-order functional does not reproduce the exact densities to a desired level of accuracy, the functional can be further improved in an iterative procedure, in which the first-order functional obtained in each iteration is considered as the known functional for the next iteration. 
\par
The model has been tested using the relativistic functional DD-PC1 as the \textit{exact} target functional. 
A simplified form of DD-PC1 has been employed for the known functional.
Assuming for the first-order correction the same functional form as in the remaining part of DD-PC1, the method described above has been applied to determine three parameters.
By employing only three finite $N=Z$ systems, and with less than ten iterations, the resulting parameters of the first-order correction are found in excellent agreement with the original parameters of the functional.
In a further illustrative calculation the target functional has been approximated by a different functional form, namely, a quadratic polynomial in the densities, determined by four parameters of the scalar and vector KS potentials.
Even though the assumed density dependence differs from that of the target functional DD-PC1, nevertheless the model accurately reproduces the density-dependent coupling functions, as well as the target densities of 
four $N=Z$ systems. 
\par
As noted in the introduction, the reason for considering relativistic functionals is that they automatically take into account the nuclear spin-orbit potential.
The inclusion of this term in the nuclear KS potential is crucial to reproduce the empirical magic numbers and shell gaps, and yet in the non-relativistic case the spin-orbit term cannot be determined by the IKS method because there is no information on the corresponding density.
The relativistic formulation does not provide a direct solution though.
The reason is that the spin-orbital potential emerges as a constructive combination of the scalar and vector nucleon potentials, but the corresponding scalar density does not represent an observable.
This brings us to the fact that accurate data exists only for charge (proton) densities, while in the IKS construction of the potential we need not only the isoscalar vector and scalar, but also the isovector densities.
A possible solution would be to combine the model developed in this work, which utilizes empirical exact densities of finite nuclei, with the equations of state of isospin symmetric and isospin asymmetric nuclear matter.
Namely, data on the proton vector densities in finite nuclei can be used together with the (microscopic) equations of state of nuclear matter to determine the isoscalar-scalar and isovector channels of the Kohn-Sham potential.
Work along these lines is in progress. 
% 
% ----------------------------------------------------------------------------------------------------------------
\begin{acknowledgments}
  This work has been supported in part by
  the Croatian Science Foundation under the project Uncertainty quantification
  within the nuclear energy density framework (IP-2018-01-5987).
  It has also been supported by the QuantiXLie Centre of Excellence, a project co-financed by the Croatian Government and European Union through the European Regional Development Fund - the Competitiveness and Cohesion Operational Programme (KK.01.1.1.01.0004),
  the RIKEN iTHEMS program, 
  the RIKEN Pioneering Project:~Evolution of Matter in the Universe,
  and
  the JSPS Grant-in-Aid for Scientific Research (Grant Nos.~18K13549, 19J20543, and 20H05648).
  % the JSPS Grant-in-Aid for Early-Career Scientists under Grant No.~18K13549,
  % the JSPS Grant-in-Aid for JSPS Fellows under Grant No.~19J20543,
  % and 
  % the JSPS Grant-in-Aid for Scientific Research (S) under Grant No.~20H05648.
\end{acknowledgments}
% ====================================================
% \newpage

    %     
\end{document}